\documentclass[aps,twocolumn,groupedaddress,showpacs]{revtex4}
\usepackage{amsmath,amscd,amssymb,epsfig}
\begin{document}
\bibliographystyle{apsrev}

\title[Non-Markovian SR]{Non-Markovian Stochastic Resonance }
\author{Igor Goychuk}
\author{Peter H\"anggi}
\affiliation{Institute of Physics, University of Augsburg,
Universit\"atsstr. 1, D-86135 Augsburg, Germany}

\date{\today}

\begin{abstract} 
The phenomenological linear response theory of non-Markovian
Stochastic Resonance (SR) is put forward for  stationary two-state
renewal processes. In terms of a derivation of a non-Markov regression
theorem we evaluate the characteristic SR-quantifiers; i.e. the
spectral power amplification (SPA) and the signal-to-noise ratio
(SNR), respectively. In clear contrast to Markovian SR, a
characteristic benchmark of genuine non-Markovian SR is its
distinctive dependence of the SPA and SNR on small (adiabatic)
driving frequencies; particularly, the adiabatic SNR becomes strongly
suppressed over its Markovian counterpart. This non-Markovian SR
theory is elucidated for a fractal gating dynamics of a potassium ion
channel possessing an infinite variance of closed sojourn times.

\end{abstract}

\pacs{05.40.-a, 82.20.Uv, 87.16.Uv}

\maketitle

The concept of Stochastic Resonance (SR), originally put forward for the description of
the periodicity of  the earth's glacial recurrences
 \cite{first}, has acquired an immense popularity in the context
of weak signal transduction in stochastic nonlinear systems \cite{review1}.
The phenomenon of SR is seemingly rather paradoxical:
an optimal dose of either external or internal noise can considerably
boost signal transduction. The archetypical situation of SR involves a
periodically rocked, continuous state  bistable
dynamics  driven by
thermal, white noise
\cite{review1}. The essential features of the perturbed bistable dynamics
can be captured by
a two-state stochastic process
$x(t)$ that switches forth and back between two metastable states $x_{1}$ and $x_{2}$
at random time points $\{ t_i\}$. This two-state random process
can be directly extracted from filtered experimental data and subsequently
statistically analyzed.

If the sojourn time intervals
$\tau_i=t_{i+1}-t_i$ are {\it independently} distributed (an assumption being invoked throughout the following),
the resulting two-state
renewal process is specified by two residence time distributions
(RTDs) $\psi_{1,2}(\tau)$ \cite{cox}. Commonly, one follows the reasoning of McNamara and Wiesenfeld \cite{mcnamara};
i.e. one approximates the reduced dynamics by
 a two-state  Markovian process with the corresponding RTDs being strictly
exponential,  $\psi_{1,2}(\tau)=\nu_{1,2}\exp(-\nu_{1,2}\tau)$. Here,
$\nu_{1,2}$ are the transition rates which are given by the inverse mean
residence times $\langle \tau_{1,2}\rangle :=\int_{0}^{\infty}
\tau\psi_{1,2}(\tau)d\tau$; i.e. $\nu_{1,2}=\langle
\tau_{1,2}\rangle^{-1}$. The input signal $f(t)$ yields time-dependent
transition rates $\nu_{1,2}\to\nu_{1,2}(t)$. The  probabilities
$p_{1,2}(t)$ of the states $x_{1,2}$ obey the Markovian master equation
\cite{review1,mcnamara}; i.e.
\begin{eqnarray}\label{rate-eq}
\dot p_1(t)& = &-\nu_1(t)p_1(t)+\nu_2(t)p_2(t)\nonumber \\
\dot p_2(t)& = &\;\;\;\nu_1(t)p_1(t)-\nu_2(t)p_2(t).
\end{eqnarray}
with the time-dependent rates.
Applying a weak periodic signal of the form
$f(t)=f_0\cos(\Omega t)$\/ yields for the asymptotic linear response
$\langle \delta x(t)\rangle=
f_0|\tilde\chi(\Omega)|\cos(\Omega t-\varphi(\Omega))$.
Here, $\tilde\chi(\Omega)$ is the linear response function in the
frequency domain and $\varphi(\Omega)$ denotes the phase shift. For
adiabatic, Arrhenius-like transition rates $\nu_{1,2}(t)$ that depend on
temperature $T$ and driving signal $f(t)$, the linear response function
$\tilde \chi(\Omega)$ is known explicitly \cite{mcnamara}. The SPA \cite
{JH91,new}, $\eta=|\tilde\chi(\Omega)|^2$, then displays the phenomenon of
SR; i.e. the quantity $\eta$ depicts a bell-shaped behavior on the thermal
noise strength  $T$ \cite{review1}. This appealing  two-state Markovian
SR-theory  due  to McNamara and Wiesenfeld \cite{mcnamara} enjoys great
popularity and wide spread application in SR-research \cite{review1}.
Moreover,  this seminal Markovian scheme has recently been generalized in
order to unify the various situations of SR  -- such as periodic or
aperiodic SR \cite{b3} and non-stationary SR -- within a unified framework
based on information theory \cite{GH00}.

One may encounter, however,  an ample number of other physical situations
where the observed stochastic two-state dynamics $x(t)$ exhibits strong
temporal long range correlations that are manifestly {\it non-Markovian}
in nature with profoundly non-exponential, experimentally observed RTDs
\cite{boguna,sullivan,mercik,teich}. In principle, any deviation of RTDs
from a strictly exponential behavior constitutes a deviation from a
Markovian two-state behavior \cite{boguna}, although in practice it can be
rather small. A clear-cut, genuine non-Markovian situation emerges 
when, e.g., one
of RTDs possesses a very large, possibly infinite variance ${\rm
var}(\tau_{1,2})=\int_{0}^{\infty} (\tau-\langle
\tau_{1,2}\rangle)^2\psi_{1,2}(\tau)d\tau\to\infty$. As a specific
example, this situation  occurs for the stochastic dynamics of the
conductance fluctuations in biological ion channels for which the RTDs
generally assume a non-exponential behavior. The corresponding RTD
$\psi(\tau)$ can either be described by a stretched exponential
\cite{sullivan}, or possibly also by a power law  $\psi(\tau) \propto
1/(b+\tau)^{\beta},\;\beta>0$ \cite{mercik}. The case with a power law is
particularly interesting: In Ref. \cite{mercik} one finds that the closed
time RTD for a large conductance (BK) potassium channel assumes a  power
law with an exponent $\beta\approx 2.24$ implying that ${\rm
var}(\tau_{closed})=\infty$. As a consequence, the conductance
fluctuations are expected to exhibit a characteristic $1/f^{\alpha}$ noise
power spectrum $S(f)$ \cite{teich}. Indeed, this result has been confirmed
for the BK ion channel \cite{siwy}, as well as for  other types of ion
channels \cite{bezrukov2000}.

What are the characteristic signatures of non-Markovian  SR in these
and several other, manifestly non-Markovian phenomena? To address
this challenge we herewith put forward the non-Markovian
generalization of the well-known McNamara-Wiesenfeld two-state Markov
theory to the case with arbitrary, non-exponential (!) RTDs
$\psi_{1,2}(\tau)$ and corresponding survival probabilities
$\Phi_{1,2}(\tau)= \int_{\tau}^{\infty}\psi_{1,2} (\tau')d\tau'$,
respectively \cite{cox}. There do exist a few prior studies of
non-Markovian SR based on a contraction of a (Markovian) stochastic
dynamics onto a non-Markovian process; see, e.g., in
\cite{review1,new,melnikov,lindner}. 
However, the case of genuine non-Markovian SR with an infinite
variance of sojourn times has not been investigated previously.
Moreover,
in clear contrast to these prior
studies \cite{new,melnikov,lindner} we do {\it not}
presume here any knowledge of the underlying microscopic or mesoscopic
dynamics. In practice, such a mesoscopic dynamics is not accessible,
or is simply not known. Instead, we pursue with this work a
phenomenological scheme of non-Markovian SR which is solely based on 
the experimentally observed RTDs $\psi_{1,2}(\tau)$ in the absence of
an input signal.

{\it Propagator for two-state renewal processes.}
A first challenge presents the derivation of the propagator
$\mathbf \Pi(t|t_0)$ of the unperturbed persistent two-state renewal
process $x(t)$. The quantity $\mathbf \Pi(t|t_0)$
relates the probability vector $\vec p(t)=[p_1(t),p_2(t)]^T$ at two different
instants of time $t$ and $t_0$, i.e.,
$\vec p(t)={\mathbf \Pi} (t|t_0)\vec p(t_0)$.
One can explicitly find $\mathbf \Pi(t|t_0)$ by considering
the various contributions of all possible stochastic paths that lead
from $\vec p(t_0)$ to  $\vec p(t)$. Let us split up
these contributions
as follows
\begin{eqnarray}\label{structure}
{\mathbf \Pi} (t|t_0)=\sum_{n=0}^{\infty}{\mathbf \Pi} ^{(n)}(t|t_0),
\end{eqnarray}
where the index $n$ denotes the number of corresponding switches that occurred during
the stochastic evolution. The contribution with zero alternations
is obviously given by
\begin{eqnarray}\label{first}
{\mathbf \Pi} ^{(0)}(t|t_0)=\left [ \begin{array}{cc}
 \Phi_1^{(0)}(t-t_0) &  0 \\
 0 & \Phi_2^{(0)}(t-t_0)
\end{array} \right ]\;.
\end{eqnarray}
Stochastic paths with a single alternation contribute
the weight
\begin{eqnarray}
{\mathbf \Pi} ^{(1)}(t|t_0)=\int_{t_0}^{t}dt_1{\mathbf P}(t-t_1){\mathbf
F}^{(0)}(t_1-t_0),
\end{eqnarray}
where
\begin{eqnarray}
{\mathbf P}(t-t_0)=\left [ \begin{array}{cc}
 \Phi_1(t-t_0) &  0 \\
 0 & \Phi_2(t-t_0)
\end{array} \right ]
\end{eqnarray}
and
\begin{eqnarray}\label{second}
{\mathbf F}^{(0)}(t-t_0)=\left [ \begin{array}{cc}
 0 & \psi_2^{(0)}(t-t_0)  \\
  \psi_1^{(0)}(t-t_0) & 0
\end{array} \right ]\;.
\end{eqnarray}
Note that for the persistent renewal process
to be strictly {\it stationary} \cite{tunaley},
the survival probability $\Phi_{1,2}^{(0)}(\tau)$ of
the {\it first} residence time interval $\tau_0=t_1-t_0$ in Eq. (\ref{first})
and the corresponding
RTD $\psi_{1,2}^{(0)}(\tau)=-d\Phi_{1,2}^{(0)}(\tau)/d\tau$ in
Eq. (\ref{second}) must be chosen differently
from all subsequent ones. Stationarity requires that \cite{cox,boguna,tunaley},
\begin{eqnarray}\label{pdf2}
\psi_{1,2}^{(0)}(\tau)=\frac{\Phi_{1,2}(\tau)}{\langle
\tau_{1,2}\rangle},
\end{eqnarray}
where $\Phi_{1,2}(\tau)=\int_{\tau}^{\infty}\psi_{1,2}(t)dt$ are the given
survival probabilities.
From (\ref{pdf2}) it follows  that the mean residence time
$\langle \tau_{1,2}\rangle$
must assume finite values, $\langle \tau_{1,2}\rangle\neq \infty$. This imposes a salient
 restriction.
Next, the  paths with two switches contribute to
Eq. (\ref{structure}) as
\begin{eqnarray}
{\mathbf \Pi}
^{(2)}(t|t_0)=\int_{t_0}^{t}dt_2\int_{t_0}^{t_2}dt_1{\mathbf
P}(t-t_2) {\mathbf F}(t_2-t_1)\\ \nonumber
\times  {\mathbf F}^{(0)}(t_1-t_0),
\end{eqnarray}
where
\begin{eqnarray}\label{endstructure}
{\mathbf F}(t-t_0)=\left [ \begin{array}{cc}
 0 & \psi_2(t-t_0)  \\
  \psi_1(t-t_0) & 0
\end{array} \right ]\;,
\end{eqnarray}
and, likewise, for all higher $n$. Because
${\mathbf \Pi} (t|t_0)$
depends only on the time difference, $\tau=t-t_0$,
the infinite,
multiple-integral  series (\ref{structure})-(\ref{endstructure})
can be summed exactly by use of a Laplace-transform. If we denote
the Laplace transform for a function $A(\tau)$ by
$\tilde A(s):=\int_{0}^{\infty}\exp(-st)A(\tau)d\tau$
we find
\begin{eqnarray}
\label{stat}
\tilde {\mathbf  \Pi}(s)=\frac{1}{s}\left [ \begin{array}{cc}
 1- \frac{\tilde G(s)}{s\langle \tau_1\rangle} &
 \frac{\tilde G(s)}{s\langle \tau_2\rangle} \\
  \frac{\tilde G(s)}{s\langle \tau_1\rangle} &
  1-\frac{\tilde G(s)}{s\langle \tau_2
  \rangle}
\end{array} \right ]\;,
\end{eqnarray}
where
\begin{equation}\label{aux}
\tilde G(s)=\frac{\left(1-\tilde \psi_1(s)\right)
\left(1-\tilde \psi_2(s)\right)}{\left(1-\tilde \psi_1(s)\tilde
\psi_2(s)\right)},
\end{equation}
in agreement with the known result
in Refs. \cite{cox,boguna}.

{\it Non-Markov Regression Theorem.} From
(\ref{stat})-(\ref{aux}) one finds the
stationary probabilities
 as $\vec p^{st}=\lim_{s\to
0}\left (s \tilde {\mathbf  \Pi}(s)\vec p(0) \right)$.  These  explicitly read
$p_{1,2}^{st}=\langle \tau_{1,2}\rangle/[\langle \tau_1\rangle+\langle
\tau_2\rangle]$.
The generally non-exponential relaxation of $\langle x(t)\rangle$ to the
stationary mean value $x_{st}=x_1p_1^{st}+x_2p_2^{st}$ is described
by the unique relaxation function $R(\tau)$, i.e.
\begin{eqnarray}\label{relax}
p_{1,2}(t_0+\tau)=p_{1,2}^{st}+[p_{1,2}(t_0)-p_{1,2}^{st}]\;
 R(\tau)
 \end{eqnarray}
where $R(\tau)$ obeys the Laplace-transform
\begin{eqnarray}\label{R-function}
\tilde R(s)=\frac{1}{s}-\left(\frac{1}{\langle \tau_1\rangle} +
\frac{1}{\langle \tau_2\rangle} \right) \frac{1}{s^2}
\tilde G(s)\;.
\end{eqnarray}

Let us consider next the
autocorrelation function
\begin{eqnarray}\label{correlation}
k(\tau)=
\lim_{t\to\infty}\frac{\langle\delta x(t+\tau)\delta x(t)\rangle }
{\langle\delta x^2\rangle_{st}}
\end{eqnarray}
of stationary fluctuations, $\delta x(t)=x(t)-x_{st}$.
With $\langle \delta x(t+\tau)\delta x(t)\rangle=
 \langle x(t+\tau) x(t)\rangle-\langle x\rangle_{st}^2$
 as $t\to \infty$,
 and
\begin{eqnarray}\label{explicit}
\lim_{t\to\infty}\langle x(t+\tau) x(t)\rangle
=\sum_{i=1,2}\sum_{j=1,2}x_ix_j\Pi_{ij}(\tau)p_j^{st}
\end{eqnarray}
we find the same result as in Ref. \cite{strat}; i.e.
\begin{eqnarray}\label{laplace-corr}
\tilde k(s)=\frac{1}{s}-\left(\frac{1}{\langle \tau_1\rangle} +
\frac{1}{\langle \tau_2\rangle}\right)\frac{1}{s^2}\tilde G(s).
\end{eqnarray}
Upon comparison of (\ref{R-function}) with (\ref{laplace-corr})
the following regression theorem holds for these non-Markovian
two-state processes, namely
\begin{equation}\label{regression}
R(\tau)=k(\tau)\;.
\end{equation}
The regression theorem (\ref{regression}), which relates the decay of 
the relaxation
function  $R(\tau)$ to the decay of stationary fluctuations
 $k(\tau)$, presents
 a first main result, yielding
 the cornerstone for the derivation
of linear response theory for non-Markovian SR.

{\it Linear Response Theory.} The common linear response approximation
  \begin{eqnarray}\label{response2}
\langle \delta x(t)\rangle:=
\langle x(t)\rangle - x_{st}
=\int_{-\infty}^{t}\chi(t-t')f(t')dt',
\end{eqnarray}
clearly holds independently of the underlying stochastic dynamics \cite{HT82}.
In (\ref{response2}),
$\chi(t)$ denotes the linear response function in the time domain. It can
be found following  an established  
procedure \cite{kubo}: (i) apply a small
static ``force'' $f_0$, (ii) let the process $x(t)$ relax to the constrained
stationary equilibrium $\langle x(f_0)\rangle$, and (iii)  suddenly remove the ``force''
at $t=t_0$. Then, in accord with (\ref{response2}) the response function reads
\begin{eqnarray}\label{response3}
\chi(\tau)=-\frac{1}{f_0}\frac{d}{d\tau}
\langle\delta x(t_0+\tau)\rangle, \;\;\;\;\;\tau>0,
\end{eqnarray}
where $\langle\delta x(t_0+\tau)\rangle=x_1p_1(t_0+\tau)+x_2p_2(t_0+\tau)$
is determined by  (\ref{relax}) with the initial $p_{1,2}(t_0)$ taken as
$p_{1,2}(t_0)=\langle \tau_{1,2}(f_0) \rangle/[\langle
\tau_{1}(f_0)\rangle+\langle \tau_{2}(f_0)\rangle]$. Expanding
$p_{1,2}(t_0)$ to first order in $f_0$ we obtain
\begin{eqnarray}\label{expand}
\langle\delta x(t_0+\tau)\rangle =
\frac{\langle \delta x^2\rangle_{st}}{\Delta x}
[\beta_2-\beta_1] R(\tau) f_0 +o(f_0),
\end{eqnarray}
where $\Delta x=x_2-x_1$ is the fluctuation amplitude and
\begin{equation}\label{msq}
\langle \delta x^2\rangle_{st}=(\Delta x)^2
\frac{\langle \tau_1\rangle\langle \tau_2\rangle}{(\langle \tau_1\rangle+
\langle \tau_2\rangle)^2},
\end{equation}
is the mean squared amplitude of the stationary fluctuations. Moreover,
$\beta_{1,2}:=d\ln \langle \tau_{1,2}(f_0)\rangle/df_0|_{f_0=0}$
in (\ref{expand}) denotes the logarithmic derivative of
mean residence time  with
respect to the input-signal strength.
Upon combining (\ref{expand})  with the regression
theorem (\ref{regression}) we obtain from (\ref{response3})
the {\it fluctuation theorem}
\begin{eqnarray}\label{fluctuation1}
\chi(\tau)=-[\beta_2-\beta_1]\frac{\theta(\tau)}{\Delta x}
\frac{d}{d\tau} \langle\delta x(t+\tau)\delta x(t)\rangle_{st} \;.
\end{eqnarray}
$\theta(t)$ denotes the unit step function. The non-Markovian fluctuation theorem
(\ref{fluctuation1}) presents a second main result of this work; in
particular, it does not assume thermal equilibrium \cite{HT82}. If, in
addition, the mean residence times commonly obey an Arrhenius-like 
dependence  on
temperature $T$ and force $f_0$; i.e.
\begin{eqnarray}\label{Arrhenius}
\langle \tau_{1,2}(f_0)\rangle & = &A_{1,2} \exp\Big (\frac{
\Delta U_{1,2}\mp \Delta x_{1,2} f_0}{k_BT}  \Big),
\end{eqnarray}
where $\Delta U_{1,2}$ are the heights of activation barriers,
$\Delta x_1=z\Delta x$, $\Delta x_2=(1-z)\Delta x$ with $0<z<1$,
we recover for the fluctuation theorem in (\ref{fluctuation1}) the
 form which in particular holds true  for a
classical  equilibrium
dynamics \cite{HT82,kubo}; i.e.,
\begin{eqnarray}\label{equilibrium}
\chi(\tau)=-\frac{\theta(\tau)}{k_BT}
\frac{d}{d\tau} \langle\delta
x(\tau)\delta x(0)\rangle_{st}.
\end{eqnarray}

{\it Spectral Power Amplification.} In presence of an applied periodic
signal, see below (\ref{rate-eq}), the  spectral power amplification (SPA) \cite{JH91},
$\eta(\Omega)=|\tilde\chi(\Omega)|^2$, where
$\tilde\chi(\omega)=\int_{-\infty}^{\infty}\chi(t){\rm e}^{i\omega t}dt$,
reads by use of the FT in (\ref{equilibrium}) upon combining
  (\ref{correlation}), (\ref{laplace-corr}),
(\ref{msq}), (\ref{Arrhenius}) as follows

\begin{equation}\label{res1}
\eta(\Omega,T)=
\frac{(\Delta x/2)^4}{(k_BT)^2}
\frac{\nu^2(T)}{\cosh^4\left[\epsilon(T)/(2k_BT)\right]}
\frac{|\tilde G(i\Omega)|^2}{\Omega^2}.
\end{equation}
In (\ref{res1}), $\nu(T)=\langle \tau_1\rangle^{-1}+\langle \tau_2\rangle^{-1}$ is the sum
of effective rates  and
$\epsilon(T)=\Delta U_2-\Delta U_1+T\Delta S$ denotes the free-energy difference
between the metastable states which includes the entropy difference
$\Delta S:=S_2-S_1=k_B\ln(A_2/A_1)$.
In the Markovian case we obtain $\tilde G(s)=s/(s+\nu)$ and  (\ref{res1}) equals
the known result, see in \cite{review1}.

{\it Signal-to-Noise Ratio.} The signal-to-noise ratio (SNR)
within linear response theory
 is given by
${\rm SNR}(\Omega,T):=\pi f_0^2|\tilde\chi(\Omega)|^2/S_N(\Omega)$,
where $S_N(\omega)$ is the spectral power of stationary fluctuations
\cite{review1}.
In the present case, $S_N(\omega)=2\langle \delta x^2 \rangle_{st}
{\rm Re}\left [\tilde k(i\omega)\right]$ with $\langle \delta x^2
\rangle_{st}$ from (\ref{msq}) and $\tilde k(s)$ given in
(\ref{laplace-corr}). By use of  (\ref{res1}), we obtain
\begin{eqnarray}\label{res3}
{\rm SNR}(\Omega,T)=\frac{\pi f_0^2(\Delta x/2)^2}{2(k_BT)^2}\frac{\nu(T)}
{\cosh^2\left[\frac{\epsilon(T)}{2k_BT)}\right]}\; N(\Omega),
\end{eqnarray}
where the term
$N(\Omega)=
|\tilde G(i\Omega)|^2/{\rm Re}[\tilde G(i\Omega)]$
denotes a frequency and temperature dependent non-Markovian correction.
For arbitrary continuous $\psi_{1,2}(\tau)$ and the high-frequency signals
$\Omega\gg \langle \tau_{1,2}\rangle^{-1}$, the 
function $N(\Omega)$ approaches
 unity.  Then, Eq. (\ref{res3}) reduces to the
 known Markovian result \cite{review1},
 i.e. the Markovian limit of SNR is assumed asymptotically in the high
 frequency limit.
More interesting, however, is the result for small frequency driving.
In the zero-frequency limit we obtain
$N(0)=2/[{\rm var}(\tau_1)/\langle \tau_1\rangle^2+
{\rm var}(\tau_2)/\langle \tau_2\rangle ^2]$.
With ${\rm var}(\tau_{1,2})=\langle \tau^2_{1,2}\rangle
- \langle \tau_{1,2}\rangle^2=
\infty$,  $N(0)=0$; i.e. ${\rm SNR}(0,T)=0$ as well.
Consequently, ultra-slow signals are
difficult to detect within the SNR-measure.

{\it Application: fractal ion channel gating.} Let us next illustrate our 
main findings
 for the case of a manifestly non-Markovian  gating dynamics of ion channels that exhibit
a fractal
gating kinetics together with a $1/f^{\alpha}$ noise spectrum of fluctuations
\cite{mercik,siwy}. In this context,
$x(t)$ corresponds to the conductance fluctuations and the forcing $f(t)$ is proportional
to the time-varying transmembrane voltage. For a locust BK channel the measured unperturbed
closed time statistics can be approximated by a Pareto law; i.e.
$\psi_1(\tau)=\langle \tau_1\rangle^{-1}(1+\gamma^{-1})/[1+\gamma^{-1}\tau/
\langle \tau_1\rangle]^{2+\gamma}$
with $\gamma\approx 0.24$ and $\langle \tau_1\rangle=0.84$ ms. The open
time RTD assumes an exponential form
with $\langle \tau_2\rangle=0.79$ ms \cite{mercik}.
For low $\omega$, the noise power reads $S_N(\omega)\propto 1/\omega^{1-\gamma}$.
Unfortunately, neither voltage, nor temperature
dependence of mean residence times  are experimentally available. Thus, we employ here the Arrhenius
dependence in (\ref{Arrhenius}).  Namely, because the
temperature dependence of the open-to-closed transitions is typically
strong \cite{hille}, we assume a rather high activation barrier; i.e.
$\Delta U_2=100$ kJ/mol
($\sim 40\;k_BT_{room}$). The closed-to-open transitions are assumed to
be  weakly temperature-dependent with $\Delta U_1=10$ kJ/mol. Because
$\langle \tau_1 \rangle \sim \langle \tau_2 \rangle$ at $T_{room}$,  the difference between
$\Delta U_1$ and $\Delta U_2$ is  compensated
by an entropy difference $\Delta S\sim
-36\;k_B$. The physical reasoning is that
the closed time statistics
exhibits a power law; i.e. the  conformations in the closed state are largely
degenerate.  This in turn yields
a larger entropy as compared to the open state.

\begin{figure}
\begin{center}
\epsfig{file=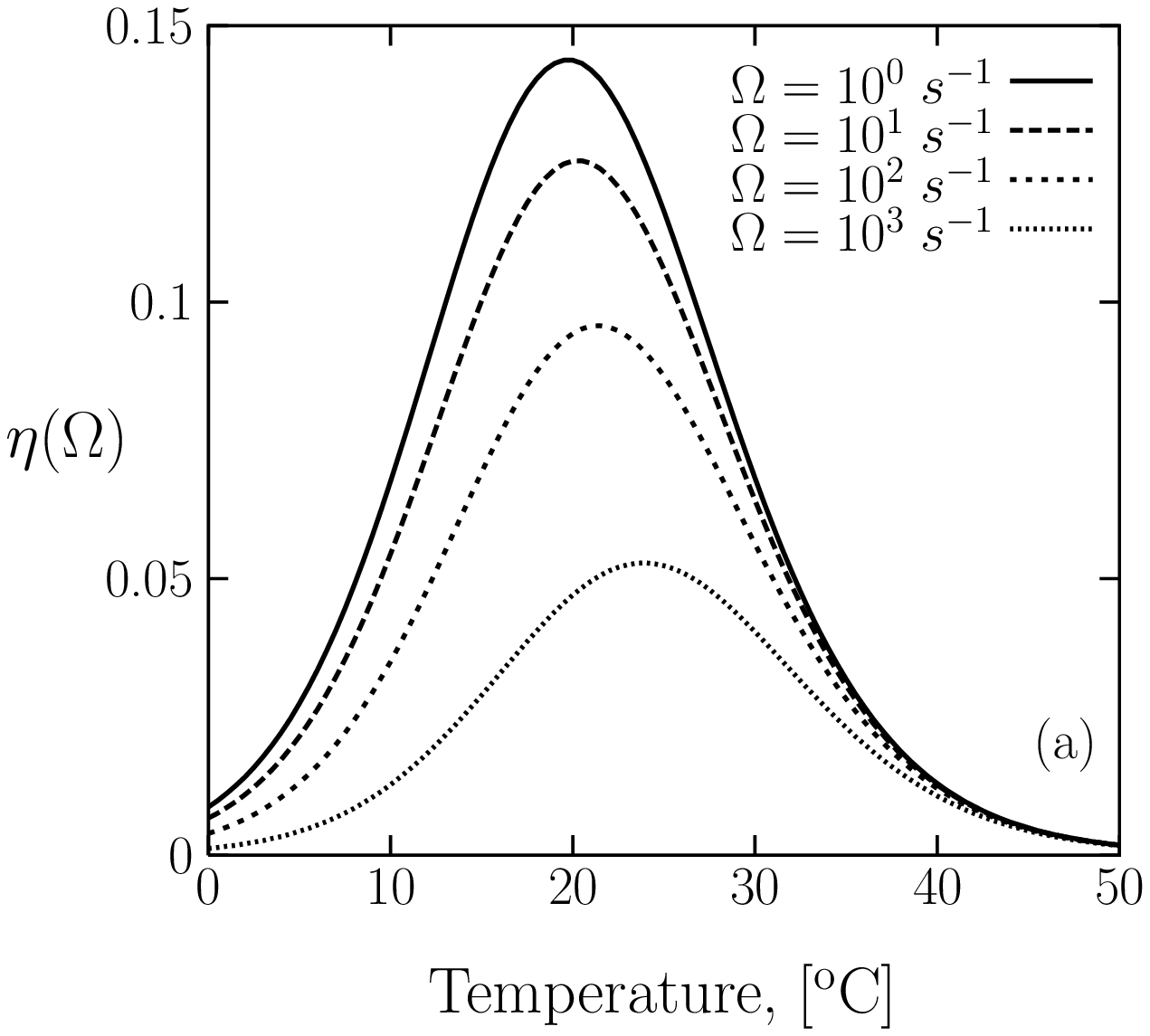,width=0.4\textwidth}
\hfill
\epsfig{file=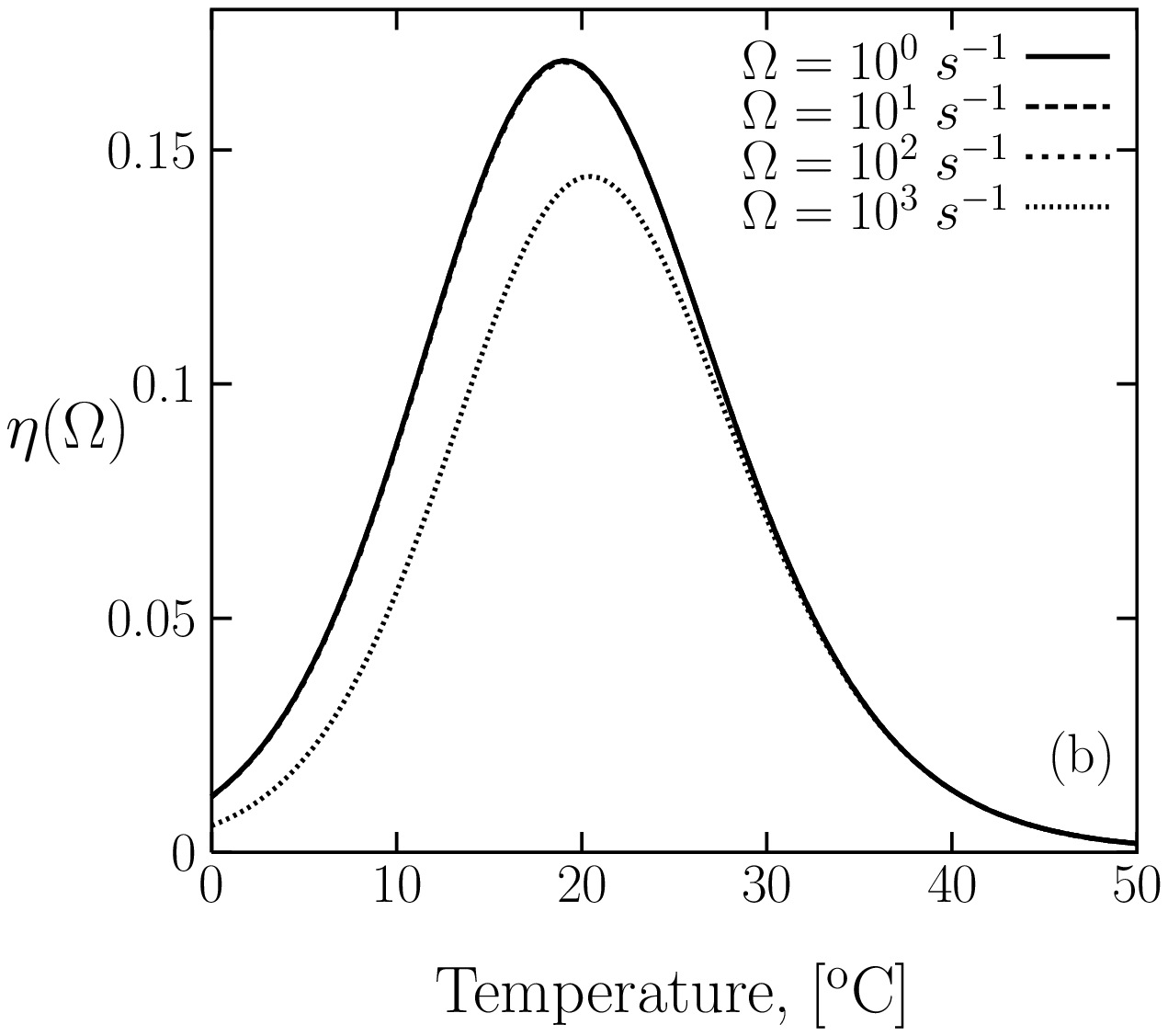,width=0.4\textwidth}
\end{center}
\caption{(a) The spectral power amplification  $\eta(\Omega)$, Eq. (\ref{res1}),
(in arbitrary units) {\it vs.} temperature (in $^o$C) for
the BK ion channel gating scenario (see text) and (b) its comparison with a corresponding Markov
modeling. }
\label{Fig1}
\end{figure}
For these parameters, the spectral power amplification  versus the
temperature
is depicted  for various driving frequencies in Fig. 1a. Furthermore, Fig. 1b
corresponds to an overall  Markovian modeling with an exponential  $\psi_1(\tau)$
possessing the same $\langle \tau_1 \rangle$. We observe a series of striking features
in Fig. 1. (i) A distinct SR-maximum occurs in
the physiological range
of varying temperatures (caused by the entropy effects). (ii) Due to
 a profound intrinsic  asymmetry, the frequency dependence of the spectral amplification
$\eta(\Omega,T)$ for the Markov modeling is very weak for small
frequencies $\Omega\ll \langle \tau_{1,2}\rangle^{-1}$ \cite{review1}. In
contrast, the non-Markovian SR exhibits a distinct low-frequency
dependence (thereby frequency-resolving the three overlapping lines
 in Fig. 1b). (iii) The evaluation of the SNR yields --
 in clear contrast to the frequency-independent Markov
 modeling -- a very strong non-Markovian SR frequency suppression of SNR
 towards smaller frequencies. The SNR-maximum
 for the top line in Fig. 1a is suppressed by two orders of magnitude
  as compared to the Markov case (not shown). As a consequence, for a
  strong non-Markovian situation
  it is preferable to use low-to-moderate frequency inputs in order to monitor SR.

In conclusion, we have put forward the phenomenological two-state
theory of non-Markovian stochastic resonance. This approach carries
great potential for many applications in physical and biological systems exhibiting
temporal long range correlations, such as they occur in life sciences
and geophysical phenomena (e.g. earthquakes), to name but
a few. In clear contrast to Markovian-SR, a benchmark of 
genuine non-Markovian SR is
its distinct  strong frequency-dependence  of  corresponding SR
quantifiers within the adiabatic driving regime; cf. Fig. 1(a).

{\it Acknowledgements.} This work has been supported by the DFG
{\it Sonderforschungsbereich} 486.

\end{document}